\documentclass[aps,prl,epsfigure,twocolumn,showpacs,floatfix]{revtex4}
\usepackage{amsmath}
\usepackage{amstext}
\usepackage{latexsym}
\usepackage{psfig}
\usepackage{graphicx}
\usepackage{amsfonts}

\newcommand{\ket}[1]{\left\vert#1\right\rangle}
\newcommand{\modul}[1]{\left\vert#1\right\vert}

\newcommand{\bra}[1]{\left\langle#1\right\vert}
\newcommand{\sand}[3]{\left\langle#1\vert#2\vert#3\right\rangle}

\begin{document}

\title{Entanglement between two superconducting qubits via interaction with non-classical radiation}
\author{ Mauro Paternostro$^\dag$, Giuseppe Falci${^\ddag}$, Myungshik Kim$^\dag$, and G. Massimo Palma$^\diamond$}
\affiliation{$^\dag$School of Mathematics and Physics, The Queen's University,
Belfast BT7 1NN, United Kingdom\\
$^\ddag$Dipartimento di Metodologie Fisiche e Chimiche (DMFCI), Universita' di Catania, viale A. Doria 6, 95125 Catania, Italy \& MATIS-INFM, Unita' di Catania\\
$^\diamond$Dipartimento di Tecnologie dell'Informazione,
Universita' di Milano,
Via Bramante 65,
26013 Crema, Italy \& NEST-INFM}
\date{\today}

\begin{abstract}
We propose a scheme to physically interface superconducting nano-circuits and quantum optics. We address the transfer of quantum information between systems having different physical natures and defined in Hilbert spaces of different dimensions. In particular, we investigate the transfer of the entanglement initially in a non-classical state of a continuous-variable system to a pair of superconducting charge qubits. This set-up is able to drive an initially separable state of the qubits into an almost pure, highly entangled state suitable for quantum information processing. 
\end{abstract}
\maketitle



Control of the dynamics of a complex quantum system requires a trade-off 
between tunability and protection against noise. To this end 
one can be interested in processes where some physical properties 
of a subsystem are reliably transferred onto the state of a second one 
(of perhaps different nature) where information can be manipulated. 
The connection between the two subsystems is effectively realized via 
a physical {\it interface}. An interface is a communication channel used to connect the elements of a quantum register to perform quantum information processing or a physical mechanism that gives full access to the system under investigation and allows to manipulate it.

\begin{figure} [ht]
{\bf (a)}\hskip3cm{\bf (b)}
\centerline{\psfig{figure=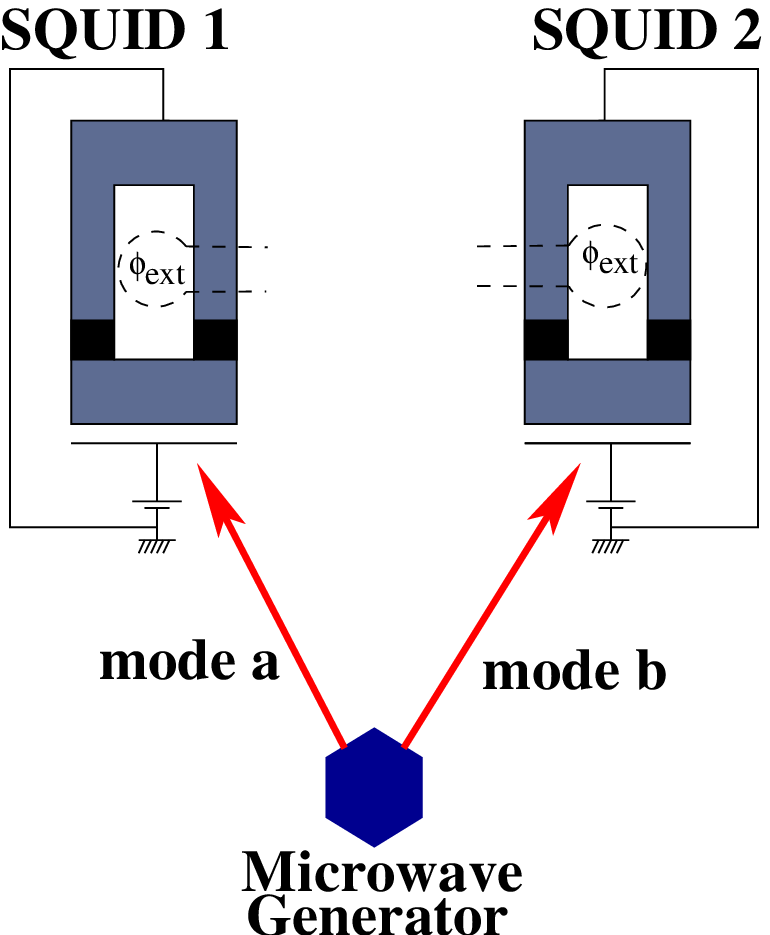,width=3.3cm,height=3.7cm}\psfig{figure=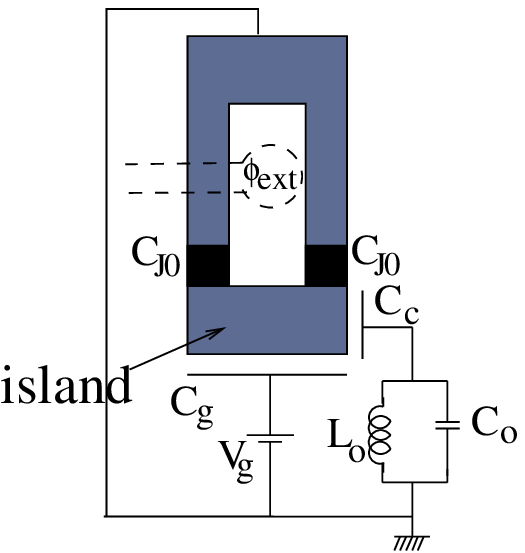,width=3.3cm,height=3.7cm}}
\caption{({\bf a}) Set-up for an entanglement transfer process via the interface between quantum correlated field modes and a pair of charge 
qubits. Each SQUID is threaded by an external magnetic flux to modulate $E_{J}(\phi_{ext})$. 
({\bf b}) Equivalent circuit for the single SQUID capacitively 
coupled to a single field mode, modeled as a LC oscillator.}
\label{mappatura}
\end{figure}

To investigate this problem, in this paper we describe the coupling between a nano-electronic circuit implementing a 
pair of quantum bits and a two-mode electromagnetic field. We discuss a mechanism for the transfer of entanglement 
from a two-mode squeezed state to the pair of qubits. Here, the information sheltered in the electromagnetic medium may be manipulated, using just single-qubit operations, when transferred to the solid-state 
subsystem. This may offer advantages with respect to integrability 
and scalability. In particular we consider a 
pair of (initially independent) Superconducting-Quantum-Interference-Devices (SQUIDs) implementing two charge qubits~\cite{Schon},
whereas each field mode is modeled as an LC circuit 
(Fig.~\ref{mappatura}). The SQUIDs can be individually addressed by gate voltages $V_{g}$ whereas an external magnetic flux $\phi_{ext}$ 
allows to change the Josephson coupling 
$E_{J}(\phi_{ext})$~\cite{Schon} and to modulate the interaction among the 
subsystems~\cite{francescopino,resonator}. Direct experimental evidence of the use of these systems as  controllable coherent two-level systems  has already been provided~\cite{exp,vion}.
 
We first analize a single SQUID plus LC oscillator (Fig.~\ref{mappatura}({\bf b})). 
We introduce the phase drop across the SQUID ($\varphi_a$) and across the 
LC circuit ($\varphi_b$). The conjugated variables 
are the excess charge on the SQUID island (${Q}_{a}$) and  the charge on 
the oscillator's capacitance ($P_{b}$). 
The Hamiltonian describing the system is:
\begin{equation}
\label{Hamiltoniana}
\begin{aligned}
&H=H_{squid}+H_{em}+H_{c}=\frac{({Q}_a - C_g V_g)^2}{2C}\\
&-E_{J}(\phi_{ext})\cos\left({\frac{2e}{\hbar}\varphi_{a}}\right) +\frac{P^2_{b}}{2C_{2}}+\frac{\varphi^2_{b}}{2L_{o}}
+\frac{P_{b}({Q}_a - C_g V_g)}{C_{1}}.
\end{aligned}
\end{equation}
where $C={\cal D}/(C_{0}+C_{c})$, $C_{1}={\cal D}/(C_{g}+2C_{J_0}+C_{c})$, $C_{2}={\cal D}/C_{c}$, ${\cal D}=(C_{o}+C_{c})(C_{g}+2C_{J_0})+C_{c}C_{o}$ and $E_{J}(\phi_{ext})=2E^{0}_{J}\cos\left(2e\phi_{ext}/\hbar\right)$.
The SQUID Hamiltonian $H_{squid}$ can be tuned by $V_g$ and $\phi_{ext}$.
The field mode, described by the oscillator $H_{em}$, has effective 
frequency $\omega=(L_{o}C_{2})^{-1/2}$ which comes from the inductance 
$L_{o}$ and the total capacitance $C_{2}$ seen by the charge $P_b$. The 
coupling Hamiltonian $H_{c}$ describes the Coulomb interaction between the charges $Q_a$ and $P_b$.



We assume large charging energy, ${e^2}/2C \gg {E}_{J}(\phi_{ext})$, and low temperatures $T \ll {e^2}/2C$. In this regime the SQUID can be described by the states $\ket{m}_{s}$ ($m=0,1$) representing $m$ Cooper pairs in excess in the island, and implements a charge Josephson qubit~\cite{Schon}.
Typical values of $C_{J_0}\simeq 10^{-15}F$ and $C_{g}\simeq10^{-17}F$ guarantee $e^2/2C\sim 1\,K{\gg}E^{0}_{J} \sim 100 \, mK$. In this system, the main sources of decoherence are noise of electrostatic origin, 
voltage fluctuations of the circuit~\cite{Schon,varenna} or stray polarization
due to charged impurities located close to the device~\cite{paladino}. 
If we set $C_{g}V_{g}={e}$, $\ket{0}_{s}$ and $\ket{1}_{s}$ have the 
same electrostatic energy and the SQUID is not affected, at first order, 
by this charge noise~\cite{vion,francescopino}. 
At this working point, 
$\hat{H}_{squid}=\frac{1}{2}E_{J}(\phi_{ext})\hat{\sigma}_{z,s}$ with a computational basis $\left\{\ket{+},\ket{-}\right\}_{s}$, where $\ket{\pm}_{s}=(1/\sqrt{2})\left(\ket{1}\pm\ket{0}\right)_{s}$ are eigenstates 
of $\hat{\sigma}_{z,s}$  splitted by $E_{J}/\hbar\sim10\,GHz$.
We introduce the operators $\hat{a}$ and $\hat{a}^\dag$ ($\left[\hat{a},\hat{a}^\dag\right]=1$) via $\hat{P}_{b}=(\hbar\omega{C}_{2}/2)^{1/2}(\hat{a}+\hat{a}^{\dagger}),\hat{\varphi}_{b}=i(\hbar/2\omega{C}_{2})^{1/2}(\hat{a}-\hat{a}^{\dag})$ and we get
$H_{em}=\hbar\omega\left(\hat{a}^{\dag}\hat{a}+1/2\right)$. Taking $C_{2}\simeq1~pF$ and $L_{o}\simeq10~nH$, achievable by present day technology, we have $\omega\simeq10~GHz$. The coupling between the SQUID and the field mode can be tuned on and off resonance 
by modulating the energy splitting of the qubit via $\phi_{ext}$. If $E_{J}(\phi_{ext})/\hbar$ is set to be much different from $\omega$, the coupling is effectively turned off and the qubit evolves independently from the field mode. On the other hand, for the quasi-resonant condition $E_{J}(\phi_{ext})\simeq\hbar\omega$, we use $\hat{Q}_{a}=2e\hat{\sigma}_{x,s}$, with 
$\hat{\sigma}_{x,s}=\left(\ket{+}_{s}\!\bra{-}\,+\,
\ket{-}_{s}\!\bra{+}\right)$,
so that
\begin{equation}
\label{accoppiamento}
\hat{H}_{c}
=
\hbar\Omega\left[(\hat{a}\hat{\sigma}_{+,s}+\hat{a}^{\dag}\hat{\sigma}_{-,s})
+(\hat{a}^{\dag}\hat{\sigma}_{+,s}+\hat{a}\hat{\sigma}_{-,s})\right],
\end{equation}
where $\Omega=e\sqrt{2\omega{C}_{2}/\hbar{C}^2_{1}}$ is the Rabi frequency of the interaction and $\hat{\sigma}_{+,s}=\hat{\sigma}^{\dag}_{-,s}=\ket{+}_{s}\!\bra{-}$.
The Hamiltonian Eq.~(\ref{accoppiamento}) is frequently found in quantum optics problems. The first and second term 
preserve the total number of excitations in the system and allow for the restriction of the computational basis in ${\cal H}_{squid}\otimes{\cal H}_{em}$ to $\left\{\ket{-,n},\ket{+,n-1}\right\}_{s,em}$.
The other (counter-rotating) terms 
induce leakage from this subspace. They can be neglected in the 
Rotating Wave Approximation (RWA) that is applicable when 
$\Omega\ll\omega,E_{J}(\phi_{ext})/\hbar$, achieved if we take 
$C_{c}\simeq10^{-17}\,F$ (weakly coupled subsystems) so that 
$C_{1}\simeq10^{-11}\,F$ and $\Omega\simeq0.1~GHz$. In this regime the 
eigenstates of SQUID+field mode are entangled 
states forming a series of doublets splitted by $\hbar\Omega\sqrt{n}$.
It is worth stressing that at the working point $V_{g}={e}/{C_{g}}$,
intra-doublet transitions are 
forbidden~\cite{francescopino}. The system is thus protected, to a certain extent, against decoherence.



\begin{figure} [ht]
\centerline{\psfig{figure=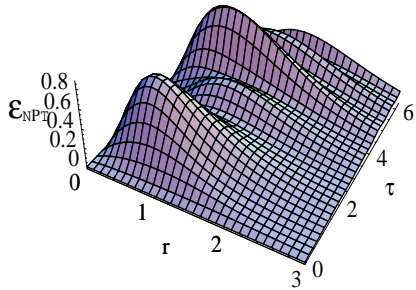,width=5.5cm,height=5.0cm}}
\caption{${\cal E}_{NPT}$ versus $\tau=\Omega{t}$ and $r$. Iff ${\cal E}_{NPT}>0$, there is entanglement between the superconducting qubits. A local maximum ${\cal E}^{max}_{NPT}\simeq0.87$ is achieved for $\tilde{r}=0.86$ and $\tilde{\tau}\simeq3\pi/2$.}
\label{negativity}
\end{figure}
\begin{figure}[b]
{\bf (a)}\hskip5cm{\bf (b)}
\centerline{\psfig{figure=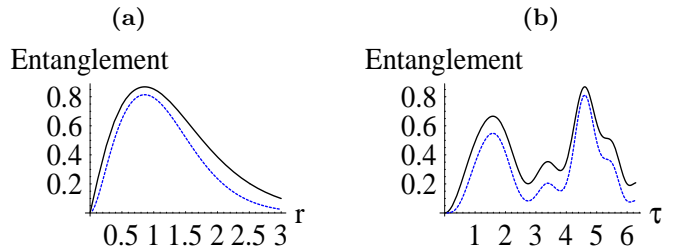,width=9.5cm,height=3.0cm}}
\vspace{0.2cm}
\caption{Comparison between EoF (dashed line) and ${\cal E}_{NPT}$ (solid line). In ({\bf a}) we plot the behavior of the two entanglement measures against the squeezing parameter $r$. We have taken $\tau=3\pi/2$. In ({\bf b}), the entanglement functions are plotted versus $\tau$, for $r=0.86$.}
\label{confrontort}
\end{figure}

We now describe the interaction of a pair of SQUIDs with non-classical radiation. We consider the two-mode squeezed state $\ket{S(r)}_{ab}=\sum^{\infty}_{n=0}\eta_{n}\ket{n,n}_{ab}$,
where $r$ is the squeezing parameter and $\eta_{n}=(\tanh{r})^{n}/\cosh{r}$~\cite{scullyzubairy}. The entanglement between modes $a,\,b$ is a function of $r$. Squeezed microwaves can be generated {\it off-line} with Josephson parametric oscillators~\cite{squeezed} and then used for our protocol. The SQUIDs can be integrated in the waveguides used for the transmission of the signal~\cite{squeezed}, with the gate-plates orthogonal to the direction of propagation of the fields. Quality factors$\,\sim{10}^4$ for a superconducting transmission-line are within the state of the art. For $\omega\sim10\,GHz$, this gives photon-lifetimes$\,\sim{1}\,\mu{sec}$ , allowing for a coherent dynamics. The SQUIDs are prepared in a pure separable state $\rho_{12}(0)=\rho_{1}(0)\otimes\rho_{2}(0)$. The interaction between each SQUID and a field mode is driven by the co-rotating part of Eq.~(\ref{accoppiamento}), $H^{ij}_{rwa}$, with $i=1,2$ and $j=a,b$. The time evolution operator is $\hat{U}(t)=\otimes_{ij}e^{-\frac{i}{\hbar}H^{ij}_{rwa}t}$~\cite{wonmin}. 
However, the reduced state of the SQUIDs $\rho_{12}(t)$ is inseparable, in general, because the evolution could have transferred quantum correlations from the fields to the qubits. To see it, we derive the operator-sum representation of the SQUIDs evolution  
$\rho_{12}(t)=Tr_{ab}\left\{\hat{U}(t)\rho_{12}(0)\otimes\rho_{ab}(0)\hat{U}^{\dagger}(t)\right\}=\sum^{}_{\mu}\sum^{\infty}_{m,p=0}\hat{K}^{mp}_{\mu}\rho_{12}(0){\hat{K}^{mp\,\dag}_{\mu}}$,
 where we the Kraus operators $\hat{K}^{mp}_{\mu}=\sum^{\infty}_{n=0}\eta_{n}(r)~\sand{m,p}{\hat{U}(t)}{n,n}$ have been introduced. Calculating the matrix elements of $\hat{U}(t)$ over the number states of the field modes, a set of five Kraus operators is found. If the initial state of the two SQUIDs is specified, a simplification is possible and the number of Kraus operators is reduced. We assume $\rho_{12}(0)=\ket{-,-}_{12}\bra{-,-}$, that can be prepared using standard techniques~\cite{Schon}. We get the effective representation $\rho_{12}(t)=\sum^{3}_{\mu=1}\sum^{\infty}_{m=0}\hat{K}^{m}_{\mu}\ket{-,-}_{12}\bra{-,-}{\hat{K}^{m\,\dag}_{\mu}}
$, where 
\begin{equation}
\label{kraus--}
\begin{split}
&\hat{K}^{m}_{1}=\eta_{m}\cos^{2}(\Omega\sqrt{m}{t})\ket{-,-}_{12}\!\bra{-,-}\\
&-\eta_{m+1}\sin^{2}(\Omega\sqrt{m+1}{t})\ket{+,+}_{12}\!\bra{-,-},\\
&\hat{K}^{m}_{2}=\eta_{m}\cos(\Omega\sqrt{m}{t})\sin(\Omega\sqrt{m}{t})\ket{-,+}_{12}\!\bra{-,-},\\
&\hat{K}^{m}_{3}=\eta_{m}\cos(\Omega\sqrt{m}{t})\sin(\Omega\sqrt{m}{t})\ket{+,-}_{12}\!\bra{-,-}.
\end{split}
\end{equation} 
$\hat{K}^{m}_{1}$ is responsible for zero and two-photon processes
 that leave the two field modes with the same number of photons. $\hat{K}^{m}_{2}$ and $\hat{K}^{m}_{3}$ describe single-photon processes in which one of the SQUIDs absorbs an incoming photon. Using Eqs.~(\ref{kraus--}), the density matrix of the SQUIDs, in the ordered basis $\left\{\ket{ij}\right\}_{12}$ ($i,j=+,-$)
, takes the form
\begin{equation}
\label{matricedensita}
\rho_{12}(r,t)=
\begin{pmatrix}
A(r,t)&0&0&-D(r,t)\\
0&B(r,t)&0&0\\
0&0&B(r,t)&0\\
-D(r,t)&0&0&C(r,t)
\end{pmatrix}
.
\end{equation} 
Here $A(r,t)=\sum^{\infty}_{n,0}\chi_{nn}(r)\cos^{4}(\Omega\sqrt{n}t),\,B(r,t)=\sum^{\infty}_{n,0}\chi_{nn}(r)\sin^{2}(\Omega\sqrt{n}t)\cos^{2}(\Omega\sqrt{n}t),\,D(r,t)=\sum^{\infty}_{n,0}\chi_{nn+1}(r)\sin^{2}(\Omega\sqrt{n+1}t)\cos^{2}(\Omega\sqrt{n}t)$ and $C(r,t)=1-2B(r,t)-A(r,t)$, with $\chi_{nm}=\eta_{n}\eta_{m}$.
To quantify the entanglement between the qubits, we choose the negativity of partial transposition (NPT). NPT is a necessary and sufficient condition for entanglement of any bipartite qubit state~\cite{Horodecki}. The corresponding entanglement measure is defined as
${\cal E}_{NPT}=-2\lambda^{-}(r,t)$,
where $\lambda^{-}(r,t)$ is the unique negative eigenvalue of the two-qubit partially transposed density matrix $\rho^{PT}_{12}$~\cite{Horodecki}.
In our case, just $\lambda^{-}(r,t)=B(r,t)-D(r,t)$ can be negative for some value of $r$ and $t$ and it is used to compute the entanglement. ${\cal E}_{NPT}$ is shown in Fig.~\ref{negativity} as a function of the degree of squeezing $r$ and the rescaled interaction time $\tau=\Omega{t}$. 
It turns out that ${\cal E}_{NPT}$ never becomes negative and only asymptotically goes to zero as $r$ is increased. Once the interaction starts, the entanglement is transferred to the qubits, {\it collapsing} and {\it reviving} as the interaction time increases. The maximum of the transferred entanglement is ${\cal E}^{max}_{NPT}=0.87$, obtained for $\tilde{\tau}\simeq3\pi/2$ and $\tilde{r}=0.86$~\cite{commento2}. ${\cal E}_{NPT}$ is not a monotone function of $r$ as can be seen in Fig.~\ref{confrontort} ({\bf a}). It is known that the correlations in $\ket{S(r)}_{ab}$ approach those of the maximally entangled EPR state when $r\rightarrow\infty$~\cite{leekim}. Increasing $r$, the contribution by higher photon-number terms in the squeezed state becomes more relevant. Each qubit is exposed to a distribution of different Rabi frequencies (each equal to $\Omega\sqrt{n}$). These induce Rabi floppings at different times that interfere spoiling the degree of entanglement between the SQUIDs. This shows that a perfectly correlated continuous variable state can not be mapped onto a maximally entangled state of two qubits. On the other hand, the discreteness of this distribution induces the entanglement to collapse and revive as time goes by. This analysis is confirmed by considering the entanglement of formation (EoF)~\cite{Wootters}. In Fig.~\ref{confrontort} the two entanglement measures are compared, as functions of both $r$ and $\tau$. From the behavior of EoF, we argue that almost one EPR singlet is required to prepare $\rho_{12}(\tilde{r},\tilde{\tau})$.

For both the measures, the SQUIDs are separable just for a short initial amount of time (${t}\simeq1\,nsec$) after which entanglement is set, persisting in time, even if fluctuating. The maximum of transferred entanglement is reached for $t\sim50\, nsec$~\cite{commento}. This is within the coherence time for the capacitive coupling considered here~\cite{francescopino} and within the lifetime of the radiation, as already stated. 
A further analysis of $\rho_{12}(\tilde{r},\tilde{\tau})$ shows that $\modul{B(\tilde{r},\tilde{\tau})}\ll\modul{A(\tilde{r},\tilde{\tau})},\modul{C(\tilde{r},\tilde{\tau})},\modul{D(\tilde{r},\tilde{\tau})}$. If, in zero-order approximation, we neglect $B(\tilde{r},\tilde{t})$ in $\rho_{12}(\tilde{r},\tilde{t})$, we get a density matrix close to that of the pure, non maximally entangled state $(\sqrt{A(r,t)}\ket{--}-\sqrt{C(r,t)}\ket{++})_{12}$. In general, $D(r,t)\neq\sqrt{A(r,t)C(r,t)}$, so that the state is mixed. The degree of mixedness in this {\it purified} version of $\rho_{12}(r,t)$ is quantified using the linearized entropy $S_{l}(\rho_{s}(r,t))=4/3\left[1-Tr\left(\rho^2_{12}(r,t)\right)\right]$,
 that ranges from 0 (pure states) to 1 (maximally mixed ones). We get $S_{l}(\rho_{12}(\tilde{r},\tilde{\tau}))\simeq0.01$ that shows that, for these parameters, the two SQUIDs are in a nearly pure state. This result is interesting: it has been proved, for example, that a bipartite mixed state becomes useless for quantum teleportation whenever its linearized entropy exceeds $1-(2/[N(N+1)])$~\cite{bosevedral}, with $N$ the dimension of each subsystem. For qubits, the threshold is $2/{3}\gg{S}_{l}(\rho_{12}(\tilde{r},\tilde{\tau}))$ and the state of the entangled SQUIDs could be used as a quantum channel in protocols for distributed quantum computation. We have calculated the purity of the state when $B(\tilde{r},\tilde{\tau})$ is included, finding the same order of magnitude of the previous result. 

\begin{figure}[ht]
{\bf (a)}\hskip4cm{\bf (b)}
\centerline{\psfig{figure=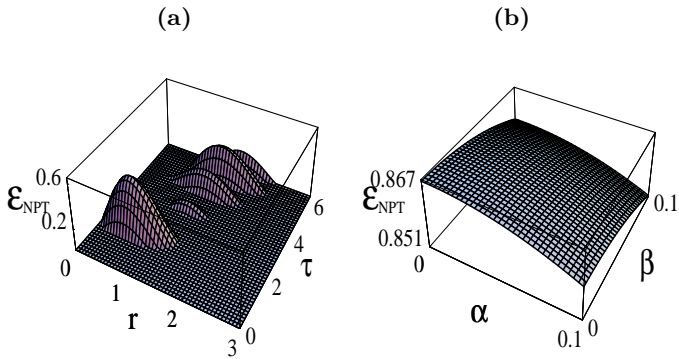,width=9.5cm,height=4.5cm}}
\caption{({\bf a}): Transferred entanglement averaged over the possible preparations of the SQUIDs. The maximum entanglement is reduced with respect to the case of $\ket{-,-}_{12}$. This is due to the contribution from $\ket{+,+}_{12}$, that is separable for wide ranges of 
$\tau$ and $r$ and spoils the average entanglement. ({\bf b}): The amount of entanglement between the qubits as a function of the 
preparation of the initial states when $r=\tilde{r}$ and 
$\tau=\tilde{\tau}$. The state $\ket{-,-}_{12}$, obtained for $\alpha=\beta=0$, corresponds to the maximum of the transferred entanglement.}
\label{entanglementmedio}
\end{figure}

We now consider the average value of the transferred entanglement as the preparation of the initial state of the SQUIDs is varied. This allows to investigate about the dynamics of the superconducting qubits once different separable states as
$\left(\cos\alpha\ket{-}+e^{i\varphi}\sin\alpha\ket{+}\right)_{1}\otimes\left(\cos\beta\ket{-}+e^{i\psi}\sin\beta\ket{+}\right)_{2}$
are considered. Here $\alpha,\beta\in\left[0,2\pi\right]$ and $\varphi,\psi\in[0,\pi]$, so that the entire Bloch sphere is explored. For simplicity, we take $\varphi=\psi=0$ and we use ${\cal E}_{NPT}$ to calculate the average value of the entanglement. The evolution of the SQUIDs involves the complete set of Kraus operators. However, the density matrix $\rho_{12}(r,t)$, averaged over an uniform distribution for $\alpha,\beta$, still keeps the form in Eq.~(\ref{matricedensita}) but with more complicated matrix elements. The results are shown in Fig.~\ref{entanglementmedio}({\bf a}). The amount of transferred entanglement is reduced and the peak at $\tau=\tilde\tau$, $r=\tilde{r}$ is shrunk to $\simeq0.4$. This can be understood considering the behavior of ${\cal E}_{NPT}$ for $\ket{+,+}_{12}$ as initial state. In this case, ${\cal E}_{NPT}$ remains negative for a wide range of values of $r$ and $\tau$ and has a small positive bump for $r\simeq0.6$ and $\tau\simeq1.7$ (that corresponds to the first peak in Fig.~\ref{negativity}). The previous result suggests that $\ket{-,-}_{12}$ plays a privileged role in the process of entanglement transfer. To support this idea, we look for the optimal preparation of the SQUIDs. 
Assuming small values of $\alpha,\,\beta$, up to the their second power, we have
\begin{equation}
\rho_{12}(r,t)\simeq{\cal N}
\begin{pmatrix}
{a}&\beta{b}+\alpha{d}&\beta{d}+\alpha{b}&-c\\
\beta{b}+\alpha{d}&a'&0&\alpha{f}-\beta{d}\\
\alpha{b}+\beta{d}&0&a'&\beta{f}-\alpha{d}\\
-c&\alpha{f}-\beta{d}&\beta{f}-\alpha{d}&a''
\end{pmatrix}
,
\end{equation}    
with ${\cal N}={(1-\alpha^{-2}-\beta^{-2})}$ and $a,a',a'',b,c,d,f$ combinations of trigonometric functions involving both $r$ and $\tau$. An analytical expression for the eigenvalues of the partial transpose is, this time, hard to be obtained. More insight is given by specifying $r$ and $\tau$. In Fig.~\ref{entanglementmedio}({\bf b}) we plot ${\cal E}_{NPT}$ versus $\alpha$ and $\beta$ for $r=\tilde{r}$ and $\tau=\tilde{\tau}$. The transferred entanglement has a maximum equal to $0.87$ for $\alpha=\beta=0$ and slowly decays. This could be important, experimentally, because small errors in the preparation of the initial state do not dramatically spoil the amount of entanglement transferred to the qubits. The same qualitative behavior is found for other values of $r$ and $t$. Thus, the initial preparation $\ket{-,-}_{12}$ provides the maximum achievable entanglement transfer. The entanglement of the SQUIDs, after the interaction, can be revealed by detecting the population $B(r,\tau)$ and the coherence $D(r,\tau)$ of the density matrix using local resonant pulses on the SQUIDs, along the same lines depicted in~\cite{sackett}.

We have proposed a physical interface between quantum optics and a system of two charge qubits. When a quantum-correlated state of light is considered, an entanglement-transfer from the field to the qubits can be efficiently tailored. This work has been supported in part by the UK Engineering and Physical Science Research Council grant GR/S14023/01. MP acknowledge IRCEP for financial support.




\begin{thebibliography}{99}

\bibitem{Schon} Y. Makhlin, G. Sch\"on, A. Shnirman, {\sl Rev. Mod. Phys.} {\bf 73}, 357 (2001); M. Tinkham, {\sl Introduction to Superconductivity} (McGraw-Hill International Editions, 1996).

\bibitem{francescopino} F. Plastina and G. Falci, {\sl Phys. Rev. B} {\bf 67}, 224514 (2003). 

\bibitem{resonator} O. Buisson and F.W.J. Hekking, in {\it Macroscopic Quantum Coherence and Quantum Computing}, D.V. Averin, B. Ruggero and P. Silvestrini Eds., Kluver (New York, 2001); F. Marquardt and C. Bruder, Phys. Rev. B {\bf 63}, 054514 (2001); 

\bibitem{exp} Y. Nakamura {\it et al.}, {\sl Nature} {\bf 398}, 786 (1999); J.M. Martinis {\it et al.}, {\sl Phys. Rev. Lett.} {\bf 89}, 117901 (2002); A. Yu. {\it et al.}, {\sl Nature} {\bf 421}, 823 (2003). 
\bibitem{vion} D. Vion {\it et al.}, {\sl Science} {\bf 296}, 886 (2002).

\bibitem{varenna} G. Falci, E. Paladino and R. Fazio, Proceedings of the International School Enrico Fermi on "Quantum Phenomena of Mesoscopic Systems", B. Altshuler and V. Tognetti Eds., IOS Bologna (2003).

\bibitem{paladino} {E.\ Paladino, L. Faoro, G. Falci and R. Fazio}, {\sl Phys. Rev. Lett.} {\bf 88}, {228304} (2002).

\bibitem{scullyzubairy} M. O. Scully and M. S. Zubairy, {\sl Quantum optics} (Cambridge University  Press, 1997). 

\bibitem{squeezed} R. Movshovich, {\it et al.}, {\sl Phy. Rev. Lett.} {\bf 65}, 1419 (1990); B. Yurke {\it et al.}, {\sl Phys. Rev. Lett.} {\bf 60}, 764 (1988); B. Yurke {\it et al.}, {\sl Phys. Rev A} {\bf 39}, 2519 (1989).

\bibitem{wonmin} W. Son, M. S. Kim, J. Lee, D. Ahn, {\sl J. Mod. Opt.} {\bf 49}, 1739 (2002).
\bibitem{Horodecki} M. Horodecki, P. Horodecki, R. Horodecki, {\sl Phys. Lett. A} {\bf 223}, 1 (1996); A. Peres, {\sl Phys. Rev. Lett.} {\bf 77}, 1413 (1996); J. Lee {\it et al.}, {\sl J. Mod. Opt.} {\bf 47}, 2151 (2000).

\bibitem{commento2} As long as correlations are present between the two modes, entanglement is set in the state of the qubits. Furthermore, we have checked that mixedness of the state of radiation due to imperfections in the generation process (squeezing thermofields instead of vacuum, for example) does not affect the entanglement transfer. 

\bibitem{leekim} M. S. Kim and J. Lee, {\sl Phys. Rev. A} {\bf 64}, 012309 (2001). 

\bibitem{Wootters} For two qubits, EoF is a monotonous function of NPT since ${\cal E}_{NPT}$ is equivalent to the {\it concurrence}. See C. H. Bennett {\it et al.}, {\sl Phys. Rev. A} {\bf 54}, 3824 (1996); W. K. Wootters, {\sl Phys. Rev. Lett.} {\bf 80}, 2245 (1998).

\bibitem{commento} We stress that the interaction between qubits and field modes is tuned on/off resonance via the external flux $\phi_{ext}$. Properly lasting {\it magnetic pulses} allow for the control of the interaction times.

\bibitem{bosevedral} This means that the usage of this state does not give a fidelity of teleportation larger than the classical one, as pointed out by S. Bose and V. Vedral, {\sl Phys. Rev. A} {\bf 61}, 040101(R) (2000). 

 
\bibitem{sackett} C. A. Sackett {\it et al.}, {\sl Nature} {\bf 404}, 256 (2000); F. Plastina, R. Fazio and G. M. Palma, {\sl Phys. Rev. B} {\bf 64}, 113306 (2001); O. Astafiev {\sl et al.} {\sl quant-ph/0402619} (2003).

\end{thebibliography}
\end{document}